\documentclass[prl,twocolumn,showpacs,superscriptaddress,aps,10pt]{revtex4-1}
\usepackage{dcolumn}
\usepackage[english]{babel}
\usepackage{graphicx}
\usepackage{amsmath,amssymb,amsthm}
\usepackage{txfonts}
\usepackage{bm}
\usepackage[utf8]{inputenc}
\usepackage{ulem}
\newcommand{\mat}[4]{\left[\begin{array}{cc} #1 & #2 \\ #3 &  #4  \end{array}\right]}

\newcommand{\ket}[1]{\left| #1 \right\rangle}

\begin{document}
\title{Programmable two-photon quantum interference in $10^3$ channels in opaque scattering media}

\author{Tom A. W. Wolterink}
\email{t.a.w.wolterink@utwente.nl}
\affiliation{Complex Photonic Systems (COPS), MESA+ Institute for Nanotechnology, University of Twente,\\ P.O. Box 217, 7500 AE Enschede, The Netherlands}
\affiliation{Laser Physics and Nonlinear Optics (LPNO), MESA+ Institute for Nanotechnology, University of Twente,\\ P.O. Box 217, 7500 AE Enschede, The Netherlands}
\author{Ravitej Uppu}
\affiliation{Complex Photonic Systems (COPS), MESA+ Institute for Nanotechnology, University of Twente,\\ P.O. Box 217, 7500 AE Enschede, The Netherlands}
\author{Georgios Ctistis}
\affiliation{Complex Photonic Systems (COPS), MESA+ Institute for Nanotechnology, University of Twente,\\ P.O. Box 217, 7500 AE Enschede, The Netherlands}
\affiliation{NanoBioInterface, Saxion University of Applied Sciences,\\ P.O. Box 70000, 7500 KB Enschede, The Netherlands}
\author{Willem L. Vos}
\affiliation{Complex Photonic Systems (COPS), MESA+ Institute for Nanotechnology, University of Twente,\\ P.O. Box 217, 7500 AE Enschede, The Netherlands}
\author{Klaus -J. Boller}
\affiliation{Laser Physics and Nonlinear Optics (LPNO), MESA+ Institute for Nanotechnology, University of Twente,\\ P.O. Box 217, 7500 AE Enschede, The Netherlands}
\author{Pepijn W. H. Pinkse}
\email{p.w.h.pinkse@utwente.nl}
\affiliation{Complex Photonic Systems (COPS), MESA+ Institute for Nanotechnology, University of Twente,\\ P.O. Box 217, 7500 AE Enschede, The Netherlands}

\date{\today}
\begin{abstract}
We investigate two-photon quantum interference in an opaque scattering medium that intrinsically supports $10^6$ transmission channels. By adaptive spatial phase-modulation of the incident wavefronts, the photons are directed at targeted speckle spots or output channels. From $10^3$ experimentally available coupled channels, we select two channels and enhance their transmission, to realize the equivalent of a fully programmable $2\times2$ beam splitter. By sending pairs of single photons from a parametric down-conversion source through the opaque scattering medium, we observe two-photon quantum interference. The programmed beam splitter need not fulfill energy conservation over the two selected output channels and hence could be non-unitary. Consequently, we have the freedom to tune the quantum interference from bunching (Hong-Ou-Mandel-like) to antibunching. Our results establish opaque scattering media as a platform for high-dimensional quantum interference that is notably relevant for boson sampling and physical-key-based authentication.
\end{abstract}


\maketitle

\begin{figure}[ht]
\includegraphics[scale=1]{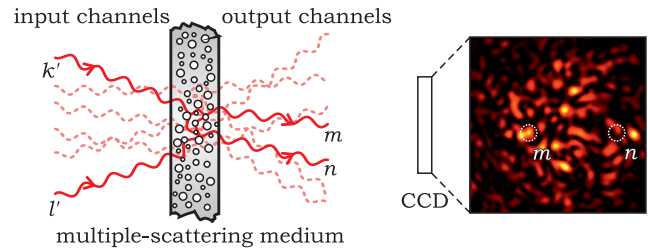}
\caption{A multiple-scattering medium couples millions of input and output channels. Light incident in the input channels results through multiple scattering in a complex interference pattern (speckle, see right panel) at the output, which can be imaged by a CCD camera. Each speckle spot in the image corresponds to an independent output channel. In this work we have programmed their interaction to create a network with 2 inputs ($k'$, $l'$) and 2 outputs ($m$, $n$).}
\label{fig:concept}
\end{figure}

Light waves propagating through an opaque scattering medium exhibit a random walk inside the medium, which is caused by multiple scattering from spatial inhomogeneities \cite{Akkermans2007}. An alternative description describes this by a transmission matrix \cite{Beenakker1997, Mosk2012}. The transmission matrix describes how a large amount of input channels is coupled to a similarly large amount of output channels, see Fig.~\ref{fig:concept}. The number of these channels can be controlled, and easily made to exceed millions, by increasing the illuminated area on the medium. Recent advances in control of light propagation through complex wavefront shaping allow for complete control over these channels in multiple-scattering media \cite{Vellekoop2007, Mosk2012, Popoff2014}. Because of their large number of controllable channels, we explore the use of multiple-scattering media to study quantum interference between multiple photons. Employed as a platform for high-dimensional quantum interference, over a large number of channels, multiple-scattering media are of relevance to boson sampling \cite{Aaronson2011, Peruzzo2010, Childs2013, Broome2013, Spring2013, Tillmann2013, Crespi2013, Spagnolo2014, Carolan2014}, quantum information processing \cite{Knill2001, Kok2007, Carolan2015, Harris2015}, and physical-key-based authentication \cite{Goorden2014}.

It has previously been observed that quantum states are robust against multiple scattering. Correlations in two-photon speckle patterns in single-scattering media have been studied \cite{Beenakker2009, Peeters2010}. Further, propagation of quantum noise \cite{Lodahl2005a, Lodahl2005b, Smolka2009} and propagation of single-photon Fock states through multiple-scattering media \cite{TJHuisman2014, Defienne2014} have also been explored. So far it has remained an open question if quantum interference of multiple photons could be demonstrated inside a multiple-scattering medium. A hurdle one might expect in an experimental implementation is the low transmission of almost all channels in the multiple-scattering medium. Remarklably, the transmission per channel is not necessarily low since complex wavefront shaping allows funneling of light into selected output modes \cite{Mosk2012, Popoff2014}.

Here we report on an experiment in which we study quantum interference in a multiple-scattering medium. We observe quantum interference of pairs of single photons in a programmable $2\times2$ beam splitter \cite{Huisman2014, Huisman2015}, made of  a multiple-scattering medium and a spatial light modulator (SLM). The spatial light modulator controls $10^3$ optical channels that are coupled in a reproducible yet unpredictable way in the multiple-scattering medium. The complex-wavefront-shaping technique using the SLM allows us to select two output channels out of $10^3$, by enhancing the amplitude of light transmitted in these channels. Programmability in this beam splitter is achieved by controlling the relative phase between the input and output arms, unlike a conventional beam splitter. We exploit this property to demonstrate not only the well-known Hong-Ou-Mandel-like bunching \cite{Hong1987}, but also the antibunching of the outgoing photon pairs, as well as any intermediate situation. Our result establishes opaque scattering media as a platform for high-dimensional quantum interference experiments as needed in, e.g., boson sampling. At present we control about $10^3$ channels, but this can readily be scaled up to a number comparable to the number of pixels in modern spatial light modulators of order $10^6$. 

\begin{figure}[ht]
\includegraphics[scale=1]{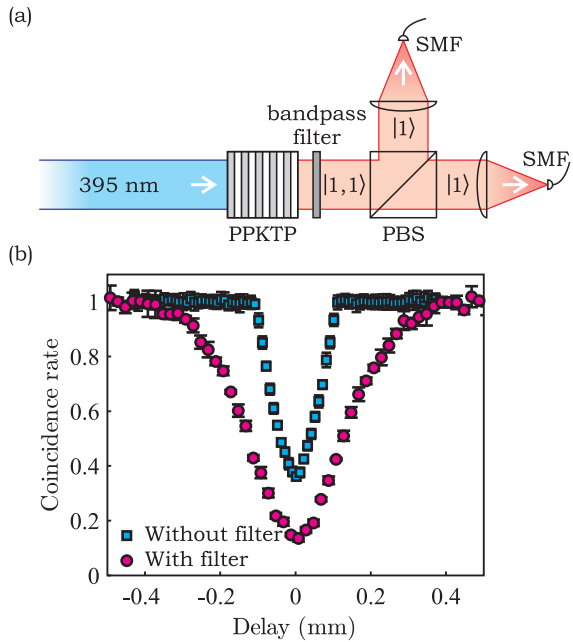}
\caption{Our quantum light source. (a) Photon pairs generated in a PPKTP crystal are separated by a polarizing beam splitter and coupled into single-mode fibers. The photons can be frequency-filtered by inserting a bandpass filter. (b) Measured Hong-Ou-Mandel interference without bandpass filter (cyan squares) and with bandpass filter (magenta circles). Error bars indicate the standard deviation in the measurements.}
\label{fig:source}
\end{figure}

Pairs of single photons in our experiment are generated using collinear Type-II spontaneous parametric down-conversion (SPDC) with a mode-locked pump at wavelength centered at 395 nm \cite{Huisman2009, Bimbard2010, TJHuisman2014} as shown in Fig.~\ref{fig:source}(a) \cite{SupplementalMaterial}. The two orthogonally-polarized single photons ($\lambda_{\rm c}$~=~790~nm) are separated using a polarizing beam splitter (PBS) and coupled into single-mode fibers (SMF). The temporal delay between the two photons can be controlled with the help of a linear delay stage in one of the single-photon channels. To measure the degree of indistinguishability of the two photons, we observe Hong-Ou-Mandel (HOM) interference at a conventional 50:50 beam splitter as shown in Fig.~\ref{fig:source}(b). At a pump power of 10~mW, a HOM dip with a visibility of 64\% is observed (cyan squares), with the usual definition of the visibility of $V\equiv\left(R_{\rm indist}-R_{\rm dist}\right)/R_{\rm dist}$, where $R_{\rm dist}$ and $R_{\rm indist}$ are the coincidence rates of distinguishable and indistinguishable photons, respectively. The visibility is less than 100\% because of some residual spectral distinguishability of the two photons that was present within the broad spectrum of the Type-II SPDC. To improve the spectral indistinguishability, a bandpass filter with a bandwidth of 1.5~nm was used that resulted in a HOM dip with an increased visibility of 86\% (magenta circles). The decrease in the spectral width also increases the width of the HOM dip as is evident in the figure.

\begin{figure}[ht]
\includegraphics[scale=1]{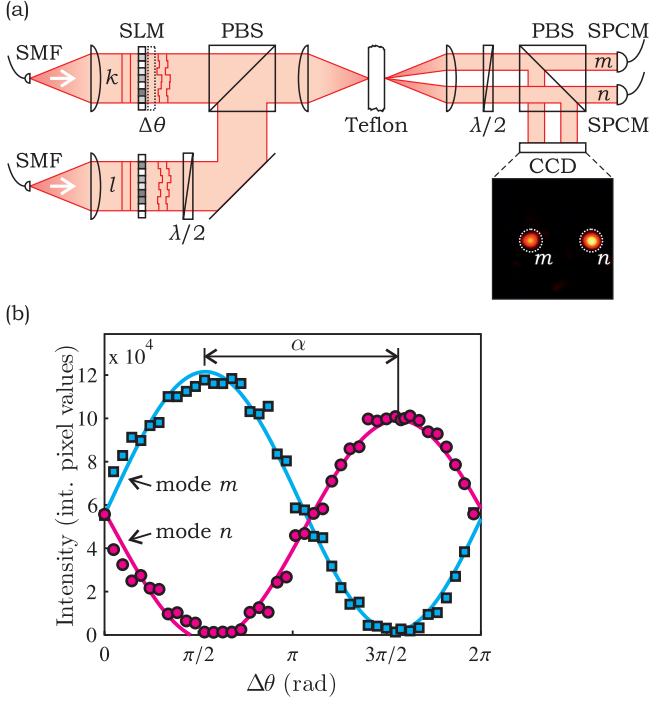}
\caption{Complex-wavefront-shaping setup for programming quantum interference. (a) Two input modes ($k$, $l$) are modulated with a SLM, spatially overlapped with orthogonal polarizations, and focused on a layer of Teflon. The transmitted light is either projected onto a CCD camera, or onto two fibers connected to SPCMs which are in this way selectively observing two output modes ($m$, $n$). (b) Interference between the input modes onto the output modes $m$ (cyan) and $n$ (magenta). The solid curves indicate sine fits to the data, resulting in $\alpha=1.02\pi$.}
\label{fig:wfs}
\end{figure}

To demonstrate programmable quantum interference in a multiple-scattering medium, we direct the light from the quantum-light source to a complex-wavefront-shaping setup with two single-mode fibers as shown in Fig.~\ref{fig:wfs}(a). The two fiber outputs have identical polarization and beam waist and form the input modes $k$ and $l$ for the quantum interference experiment. Both modes are phase modulated with a liquid-crystal spatial light modulator, and afterwards spatially overlapped using a half-wave plate and a polarizing beam splitter, resulting in a collinear propagation of the two modes with orthogonal polarizations. An objective (NA=0.95) is used to focus the light onto the scattering medium slab. The scattering medium consists of a layer of polytetrafluoroethylene (PTFE, Teflon) with a thickness of approximately 500~$\mu$m and has a scattering mean free path of 150~$\pm$~10~$\mu$m, that we determined by a coherent backscattering cone measurement. The scattering medium was found to be more stable over extended periods of time in our laboratory environment than powder-based samples, which we attribute to its impermeability and therefore insensitivity to humidity changes. The experimental setup, including multiple-scattering medium, is interferometrically stable for a duration longer than a month. The transmitted light is collected by a second objective (NA=0.6) and after transmission through a PBS it is coupled into two multimode fibers (output modes $m$ and $n$) connected to single-photon counting modules (SPCM). The multimode fibers have a core diameter of 200~$\mu$m, which is smaller than the size of a single speckle in the transmitted light, to ensure that they each collect light of only a single mode. The total number of contributing modes (speckles) is approximately $4\times10^3$. By rotating a half-wave plate this light can also be reflected off the PBS and directly projected onto a CCD camera. 

Realization of a programmable 2$\times$2 beam splitter inside a scattering medium follows a complex-wavefront-shaping process. To ease this process, we couple classical laser light ($\lambda_{\rm c}$~=~790~nm) into the single-mode fibers and monitor the light in the output modes $m$ and $n$ using amplified photodiodes. In short, the process starts with a single input mode $k$ incident on the scattering medium. We optimize the output mode $m$ by fitting the optimal phase for each SLM segment that results in maximum constructive interference in the multimode fiber \cite{Vellekoop2007}, using direct (classical) light from the laser. Each input mode is controlled by approximately 960 segments on the SLM. The intensity of the single enhanced output spot coupled to the multimode fiber is about $200$ times greater than the average intensity of the background speckles. Next we optimize the second output mode $n$ in the same way. A camera image of two optimized output modes is shown in Fig.~\ref{fig:wfs}(a). This optimization procedure is repeated for the input mode $l$, resulting in four phase patterns. By combining these four interference patterns we program any $2\times2$ circuit \cite{Huisman2015}. The transmission through the circuit considered in this experiment is described by the following equation that relates the electric field of the two output modes, $E_{m}$ and $E_{n}$, to the electric fields of the two input modes, $E_{k}$ and $E_{l}$, by means of the transmission matrix $\bf{T}$:
\begin{equation}\label{eq:matrix}
\left[\begin{array}{c} 
E_{m} \\
E_{n} \\
\end{array}\right]
= {\bf T} \left[\begin{array}{c} 
E_{k} \\
E_{l} \\
\end{array}\right]
= t \mat{1}{1}{1}{\exp{(\rm{i} \alpha)}} 
\left[\begin{array}{c} 
E_{k} \\
E_{l} \\
\end{array}\right],
\end{equation}
where the parameter $\alpha$ is set in the algorithm when combining the interference patterns. The amplitude transmission coefficient $t$ has $\lvert{t\rvert}<\frac{1}{\sqrt{2}}$, emphasizing that the circuit is inherently lossy. Since the two selected channels stem from a manifold of $10^3$ channels, the programmed $2\times2$ circuit need not fulfill energy conservation and could thus be non-unitary. Consequently, we have the freedom to tune the quantum interference from bunching (Hong-Ou-Mandel-like) to antibunching. Only for $\alpha=\pi$ and $\lvert{t\rvert}=\frac{1}{\sqrt{2}}$ we do have a unitary matrix with ${\bf T}^\dagger {\bf T}=1$, and this transmission matrix represents an ideal 50:50 beam splitter. To confirm the functionality of each circuit, classical light is injected into both input modes $k$ and $l$ and the intensities of the output modes $m$ and $n$ are monitored while applying a phase difference $\Delta\theta$ between the input modes. An example of such an interference measurement is shown in Fig.~\ref{fig:wfs}(b) for $\alpha=\pi$. A similar measurement at $\alpha=0$ shows two overlapping $\left(1+\sin{\Delta\theta}\right)$-shaped curves, indicating the inherent non-unitary behavior of this $2\times2$ circuit. After programming the functionality, we switch back from using the direct classical light of the laser to the down-converted photons from the quantum light source.

\begin{figure}[ht]
\includegraphics[scale=1]{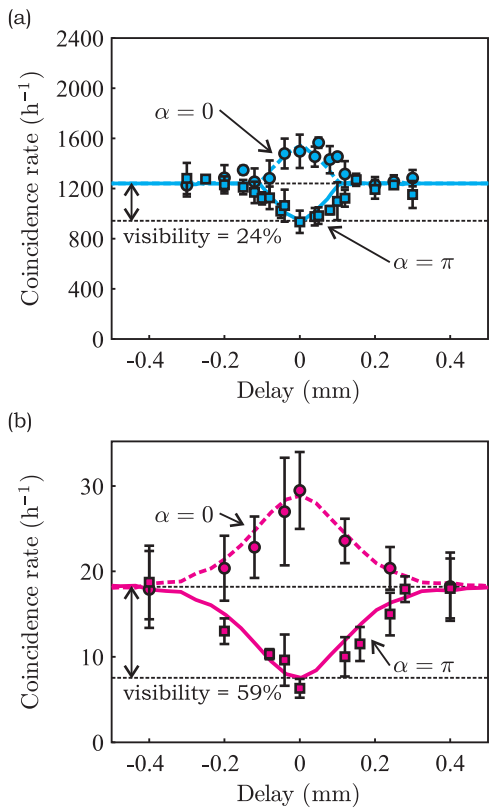}
\caption{Programmable quantum interference. Setting $\alpha=\pi$ gives rise to a dip in coincidence counts (squares), while $\alpha=0$ gives a peak (circles). Error bars indicate the standard deviations in the measurements. (a) depicts measurements performed at a pump power of 100~mW without bandpass filter, and (b) at a pump power of 40~mW with the filter in place. The solid curves in the plots correspond to the measured coincidence dips using a conventional beam splitter. The flipped traces of these are shown by the dashed curves as a guide to the eye.}
\label{fig:dips}
\end{figure}

Writing the input of the circuit described by Eq.~\eqref{eq:matrix} as $\Psi_{\rm in}=\ket{1}_{k} \ket{1}_{l}$, one can find the probabilities for all possible outputs in a straightforward manner \cite{Barnett1998} for indistinguishable photons:
\begin{equation}\label{eq:probindist}
\begin{split}
P\left(2_{m},0_{n}\right) &= P\left(0_{m},2_{n}\right) = 2{\lvert t\rvert}^4 \\
P\left(1_{m},1_{n}\right) &= 2{\lvert t\rvert}^4 \left(1+\cos{\alpha}\right) \\
P\left(1_{m},0_{n}\right) &= P\left(0_{m},1_{n}\right) = 2{\lvert t\rvert}^2 - 2{\lvert t\rvert}^4 \left(3+\cos{\alpha}\right) \\
P\left(0_{m},0_{n}\right) &= 1 - 4{\lvert t\rvert}^2 + 2{\lvert t\rvert}^4 \left(3+\cos{\alpha}\right)
\end{split}
\end{equation}
Of interest are the first three probabilities $P\left(2_{m},0_{n}\right)$, $P\left(0_{m},2_{n}\right)$, and $P\left(1_{m},1_{n}\right)$, for which both photons arrive in output modes $m$ and $n$. Using the circuit of Fig.~\ref{fig:wfs}(b) programmed with $\alpha=\pi$, representing a 50:50 beam splitter, should result in Hong-Ou-Mandel interference in the scattering medium. Figure~\ref{fig:dips}(a) shows the measured quantum interference for this circuit which indeed shows a HOM dip (squares). These measurements were done using a pump power of 100~mW and without a bandpass filter. While this higher pump power (in comparison to Fig.~1) generates more photon pairs ($> 4\times10^6$ s$^{-1}$), it also increases the production rate of higher photon-number states. These states reduce the visibility of the HOM dip. The measured HOM dip with a conventional beam splitter at this pump power indeed shows a reduced visibility of 24\% (solid curve). To consolidate the quantum nature of the interference \cite{Ghosh1987}, we repeated the measurements at a reduced pump power of 40~mW with a bandpass filter in place, which results in a visibility of 59\% as shown in Fig~\ref{fig:dips}(b) (squares). Indeed the data match the prediction and therefore confirms the quantum nature of the interference.

We now explore the programmability of the quantum interference. For indistinguishable single photons and for arbitrary $\alpha$, we expect the probability to detect coincidences between output modes $m$ and $n$ to scale as $\left(1+\cos{\alpha}\right)$. For instance, setting $\alpha=0$ in the patterns on the SLM would double the rate of coincidences compared to the rate obtained for distinguishable photons. This appears as a peak instead of a dip in the measured coincidences as shown in Fig.~\ref{fig:dips} (circles). Although the case with $\alpha$ = 0 can not be realized in a conventional beam splitter, we show a flipped trace of the measured HOM dip with a conventional beam splitter as a guide to the eye (dashed curve). Note that the probabilities for bunched photons in the outputs is independent of $\alpha$, as is evident from Eq.~\eqref{eq:probindist}, which makes the output state different from the states typically found when recombining the two outputs from a traditional HOM experiment in a Mach-Zehnder type interferometer \cite{Rarity1990, Chen2007, Silverstone2014}.

\begin{figure}[ht]
\includegraphics[scale=1]{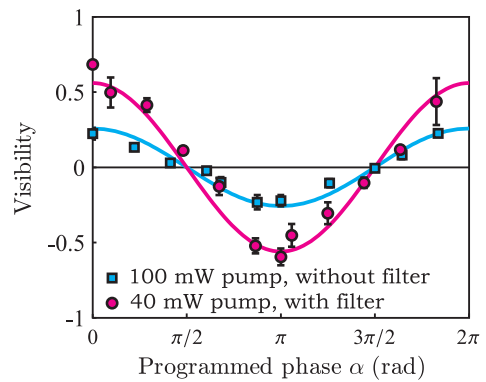}
\caption{Programmable quantum interference as a function of programmed phase $\alpha$ at a pump power of 100~mW without filter (cyan), and at 40~mW with bandpass filter (magenta). Error bars indicate the standard deviation in the measurements. The solid curves indicate $V_0\cos{\alpha}$ fits to these data with as only free parameter the prefactor $V_0$.}
\label{fig:phasescan}
\end{figure}

Fully programmable quantum interference is demonstrated in our complex scattering medium in Fig.~\ref{fig:phasescan}. Here the visibility of the quantum interference $V_0$ is plotted as a function of phase $\alpha$. Negative visibility corresponds to a dip in the coincidence counts and postive visilibty to a peak. At $\alpha=0$ a coincidence peak is observed. The visibility of this peak vanishes at phase $\alpha=\pi/2$, after which the visibility increases again as a coincidence dip. The well-known HOM dip with a high visibility occurs at phase $\alpha=\pi$. The solid curves indicate $V_0 \cos \alpha$ fits to these data with the prefactor $V_0$ as the only free parameter .

For the measurements at a pump power of 100~mW the average coincidence detection rate is only 1240~h$^{-1}$, corresponding to a single photon detection rate of 2500 s$^{-1}$. This detection rate is small in comparison to the generation rate of $17 \times 10^6$~s$^{-1}$ as a result of our choice to be deep in the multiple-scattering regime. In this way, we approach the assumption of maximal entropy of the matrix in random-matrix theory \cite{Beenakker1997}. Multimode fibers that were recently exploited for two-photon quantum interference \cite{Defienne2015}, do not provide such a high entropy, since the transmission matrix of a multimode fiber can be transformed into a block-diagonal matrix by a suitable basis \cite{Carpenter2014, Cismar2015}. We note that having maximal entropy is essential for the application of a scattering medium in quantum-secure authentication \cite{Goorden2014}, which requires a physical unclonable function with a transmission matrix that cannot predictably be approximated by a near-diagonal matrix \cite{Pappu2002}. 

In summary, we demonstrated two-photon quantum interference in a high-dimensional linear optical network consisting of an opaque scattering medium. Out of the approximately $10^6$ available channels, we control about $10^3$ channels in this linear network from which we construct a programmable $2\times2$ circuit in which we have studied quantum interference between two single photons. We have demonstrated that by programming the functionality of this circuit, the well-known Hong-Ou-Mandel bunching can be made to vanish, or be transformed into antibunching. Our results demonstrate the feasibility of using complex scattering media as high-dimensional linear optical network for quantum information processing \cite{Knill2001, Kok2007} and boson sampling \cite{Peruzzo2010, Childs2013, Broome2013, Spring2013, Tillmann2013, Crespi2013, Spagnolo2014, Carolan2014}. By using complex wavefront shaping one can achieve a programmable functionality to realize universal quantum gates \cite{Carolan2015, Harris2015}.

We thank Simon Huisman, Thomas Huisman, Ad Lagendijk, Allard Mosk, Boris \u{S}kori\'{c}, and Tristan Tentrup for discussions and support. This work is financially supported by FOM, NWO (Vici), and STW.

\end{document}